\documentclass[referee]{raa}          % referee version: for submission

\usepackage{graphicx,times}             %for PS/EPS graphics inclusion, new
\usepackage{natbib}
\usepackage{amssymb,amsmath}
\usepackage{multirow}
\bibpunct{(}{)}{;}{a}{}{,}

\usepackage[a4paper=true,dvipdfm=true,pagebackref=true]{hyperref}
\hypersetup{colorlinks = true, linkcolor = green, anchorcolor = red, citecolor = blue, filecolor = red, pagecolor = red, urlcolor = red}

\begin{document}

   \title{Spectral properties of the surface reflectance of the northern polar region of Mercury
%\,$^*$
%\footnotetext{$*$ Supported by the National Natural Science Foundation of China.}
}
%   \subtitle{I. Place Your Subtitle Here}

   \volnopage{Vol.0 (20xx) No.0, 000--000}      %%preserved for Editor. DOn't remove!
   \setcounter{page}{1}          %%starting page, preserved for Editor. DOn't remove!

   \author{Nguyen Bich Ngoc
      \inst{1}
   \and Nicolas Bott
      \inst{2}
   \and Pham Ngoc Diep
      \inst{1}
   }

   \institute{Department of Astrophysics, Vietnam National Space Center, Vietnam Academy of Science and Technology, 18 Hoang Quoc Viet, Cau Giay, Hanoi, Vietnam; {\it pndiep@vnsc.org.vn}\\
        \and
             Laboratoire d'Etudes Spatiales et d'Instrumentation en Astrophysique, l'Observatoire de Paris, 5 place Jules Janssen 92195 Meudon, France\\
\vs\no
   {\small Received~~20xx month day; accepted~~20xx~~month day}}

\abstract{We analyse MESSENGER reflectance measurements covering the northern polar region of Mercury, the least studied region of the northern mercurian hemisphere. We use observations from the Mercury Dual Imaging System Wide-Angle Camera (MDIS/WAC) and the Mercury Atmospheric and Surface Composition Spectrometer (MASCS/VIRS) to study the spectral dependence of the surface reflectance. The results obtained from the observations made by both instruments are remarkably consistent. We find that a second degree polynomial description of the measured reflectance spectra gives very good fits to the data and that the information that they carry can best be characterized by two parameters, the mean reflectance and the mean relative spectral slope, averaged over the explored range of wavelengths. The properties of the four main types of terrains known to form Mercury’s regolith in the northern region, smooth plains (SP), heavily cratered terrain (HCT), fresh ejecta/materials and red pitted ground (RPG) are examined in terms of these two parameters. The results are compared, and found consistent with those obtained by earlier studies in spite of difficulties met in obtaining accurate reflectance measurements under the large incidence angle condition characteristic of polar regions. These results will help with the preparation of the BepiColombo mission and with supporting its observational strategy.
}

   \authorrunning{N. B. Ngoc, N. Bott \& P. N. Diep}            %author_head in even pages
   \titlerunning{Spectral properties of surface reflectance}  % title_head in odd pages

   \maketitle
%
%________________________________________________ sections below
%
\section{Introduction}           %% first-level sections will be auto-capitalized
\label{sect:intro}
\subsection{Mercury Surface}
Mercury is the innermost and smallest planet of the Solar System, with a 2,440 km radius and an elliptical heliocentric orbit (between 0.31 AU at perihelion and 0.47 AU at aphelion). The planet offers a key to answer questions about formation and evolution of terrestrial planets. The study of its surface helps with the understanding of its thermal evolution as well as of its volcanic and geological history. As a result, the mechanisms governing the formation of a planet close to its host star, as well as the formation of the Moon, can be better constrained. The first Mercury mission, Mariner 10, lasted from 1973 to 1975. With three flybys and over 2,000 useful images, Mariner 10 was able to map $\sim$45\% of Mercury’s surface; it identified two main types of terrains: smooth plains (SPs) and heavily cratered terrains (HCTs). MESSENGER (MErcury Surface, Space ENvironment, GEochemistry, and Ranging) was the second mission sent to Mercury (\citealt{solomon+etal+2018}). Launched in August 2004, the spacecraft entered in orbit in March 2011, and became the first spacecraft orbiting Mercury. It crashed on the planet on April 30, 2015. It carried several instruments for the study of Mercury’s surface, two of which, the Mercury Dual Imaging System (MDIS) (\citealt{hawkins+etal+2007,hawkins+etal+2009}) and the Mercury Atmospheric Surface Composition Spectrometer (MASCS) (\citealt{mcclintock2007mercury}) are used in the present work. MESSENGER obtained the first complete picture of Mercury: global geology, surface composition, distribution of volcanism, detection of water ice on polar areas, etc. Reference (\citealt{solomon+etal+2018}) offers an exhaustive and comprehensive collection of articles summarizing our current knowledge of the planet in the wake of the MESSENGER mission. Of particular relevance to the present work is the chapter on the spectral reflectance of Mercury’s surface (\citealt{murchie2018spectral}) from which a complete list of pertinent references can be traced. MESSENGER images show SPs covering $\sim$27\% of the surface of the planet; they are younger, with a lower crater density, than HCTs. They are formed as volcanic lava flows or as ejecta deposits from basin-forming impact events (\citealt{blewett2009multispectral,denevi2009evolution,denevi2013distribution}). Volcanism, an important process in Mercury’s geologic history, is of two types: effusive volcanism associated with the SPs, and explosive volcanism with pyroclastic deposits (\citealt{goudge2014global}). Most of the SPs, more than $\sim$65\%, have probably volcanic origin and are interpreted as products of effusive volcanism (\citealt{murchie2015orbital}). HCTs have a high impact crater density suggesting that these terrains recorded the period of late heavy bombardment which ended about 3.8 billion years ago on the Moon (\citealt{solomon+etal+2001}). Compared with the Moon, Mercury’s global surface has a lower albedo and a generally steeper spectral slope with no strong mineral absorption (\citealt{robinson2008reflectance,nittler2011major,izenberg2014low}). Thanks to MESSENGER, so-called ‘hollows’ were discovered, which have not been found on any other rocky planet of the Solar System. Hollows are made of bright, fresh and spectrally immature material; they have irregular shapes and rounded edges; they are found in impact craters, both on the central peak and on the ring at the boundary (\citealt{blewett2016analysis}).

%% Authors can give a citation as 'Michel et al. 1992'.
%% You may also use \cite, \citep and \citet for citation, and use Table~1 or Figure~1
%% and so forth. Using \ref and \label for cross-references of Tables/Figures
%% is a good way in adjusting/adding/removing text, tables or figures.
%-------------------------------------------------------------------------

\subsection{Borealis Quadrangle}
The Borealis (H-01) quadrangle of Mercury, hereafter referred to simply as Borealis, covers the north pole at latitudes in excess of 65$^\circ$. With Mariner 10 mission, only $\sim$40\% of the Borealis region had been mapped. We had accordingly incomplete knowledge of the north pole for the forty years that followed. With MESSENGER observations, Borealis was fully mapped for the first time using MDIS monochrome mosaic (\citealt{blewett2009multispectral,ostrach2017geologic}). The morphology of Borealis was found (\citealt{ostrach2015extent}) to be dominated by volcanic plains, referred to as northern SPs, and by northern HCTs. The northern SPs are relatively flat and cover $\sim$2/3 of the surface of Borealis between longitudes of $\sim$130$^\circ$E and $\sim$105$^\circ$W. Current models assume that the northern SPs were formed rapidly by large volumes of low viscosity lava at high temperature (\citealt{ostrach2015extent}). The observed distribution of pyroclastic deposits of Mercury (\citealt{goudge2014global}) shows none in Borealis, excluding explosive volcanism. Many wrinkle ridges, which are common physiographic features of SPs on terrestrial planets, have been mapped in detail (\citealt{crane2019tectonic}) and several hollows have been identified at low Borealis latitudes together with previously unidentified deposits now identified as relatively red deposit and referred to as Red Pitted Ground (RPG) (\citealt{thomas2014hollows}). Crater rays, streams of materials ejected from craters, can be seen on Borealis as coming from the young Hokusai impact crater (\citealt{xiao2016self}).

%-------------------------------------------------------------------------   
\subsection{Aim of the Study}
Succeeding MESSENGER, BepiColombo, which was launched on 19$^{\textrm{th}}$ October 2018, will perform further studies of Mercury; it is the first ESA-JAXA mission dedicated to the planet (\citealt{benkhoff2010bepicolombo}). The BepiColombo Mercury Planetary Orbiter (MPO) carries imaging instruments such as SIMBIO-SYS (Spectrometer and Imagers for MPO BepiColombo Integrated Observatory SYStem) (\citealt{flamini2010simbio}) aimed at improving our knowledge of Mercury’s surface. To prepare for BepiColombo mission and optimise the outcome of its future observations, we need to extract as much information as possible from the MESSENGER data. 

The aim of the present study is to contribute additional information on the identification of the main characteristics of the surface of Borealis and the correlation of the spectral reflectance properties with the different geological units, using two sets of MESSENGER data, one from MDIS and the other from MASCS.

%-------------------------------------------------------------------------
%-------------------------------------------------------------------------
\section{Data Sets}
\label{sect:Data}
MDIS and MASCS measure the radiance of sunlight scattered by the surface of Mercury, from which the spectral reflectance is evaluated. For a Lambertian surface illuminated normally by the Sun, one expects a spectral radiance at wavelength $\lambda$ equal to the effective irradiance of the Sun on Earth, ${F_{\lambda}}$, divided by $\pi\,D^2$,  ${L_{0\lambda}=F_{\lambda}/(\pi\,D^2)}$, where the factor $\pi$ accounts for Lambertian reflection and where $D$ is the distance of Mercury to the Sun measured in astronomical units. The spectral reflectance $R_{\lambda}$ is defined as the ratio between the measured spectral radiance $L_{\lambda}$ and the value of ${L_{0\lambda}}$ taken as reference, $R_{\lambda}=L_{\lambda}/L_{0\lambda}$. The raw MDIS and MASCS data need to be processed in order to obtain maps of the spectral reflectance. In the MASCS case, data reduction, including radiometric calibration and photometric correction, was done by the MESSENGER team (\citealt{izenberg2014low, besse2015spectroscopic}). In the MDIS case, data reduction was done by us applying the procedure for calibration, photometric standardization, and processing of images described by \citealt{deveni+etal+2018} using the Integrated Software for Imagers and Spectrometers package (ISIS)\footnote{https://isis.astrogeology.usgs.gov}. 

Converting the measured radiance to a standard configuration uses the values of the incidence angle $i$, emergence angle $e$ and phase angle $\varphi$, defined as the angle between incident and emitted light, as well as reference laboratory data obtained with incidence angle $i_{ref}$, emergence angle $e_{ref}$, and phase angle $\varphi_{ref}$. 

Compared with other quadrangles, the observational orbit of MESSENGER over Borealis provides a larger number of images but requires larger values of the phase angles, in excess of 78$^\circ$ in the MASCS case, and larger values of the incidence and reflection angles, implying important distortion. It also implies, because of the low altitude of the satellite, narrower footprints. Distortion of observed images at high incidence angles is an important effect which is caused by several reasons: difficulties in co-registering for mosaicking and in modelling photometric reflectance behavior; taking into account the systematic reddening at different wavelengths (\citealt{domingue+etal+2015}, \citealt{murchie2015orbital}). In both MASCS and MDIS observations, we reject data having an incidence angle $i$ larger than 80$^\circ$.

%-------------------------------------------------------------------------
\subsection{MDIS Data}
\label{section2.1}
During the four years of MESSENGER operation, MDIS took over 42,000 images of Borealis. MDIS consists of a multispectral Wide-Angle Camera (WAC) equipped with twelve filters (seven in the visible and five in the near-infrared) and a Narrow-Angle Camera (NAC) that takes high resolution monochrome images. Here we only use WAC data; WAC has a 10.5$^\circ$ field of view covered by a CCD array of 1024$\times$1024 pixels. In the present study, we use raw WAC data with eight filters (at 433, 480, 559, 629, 749, 828, 899 and 996 nm). Another filter (700 nm) is used for radiometric calibration and the other three filters (699, 947, 1013 nm) are not used because of too small a number of images taken (less than 1\% in comparison with other filters).

\begin{figure}
   \centering
   \includegraphics[width=14cm]{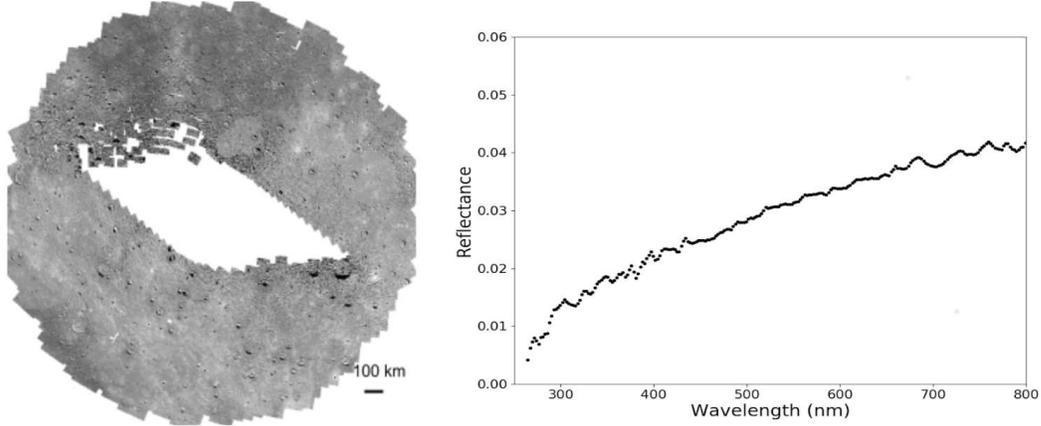}
   \caption{Left: grayscale MDIS-WAC mosaic image of Borealis at 996 nm. The blank region in the centre of the image is due to the lack of data and to the 80$^\circ$ cut on incidence angle. Right: a typical MASCS spectrum of Borealis SPs (footprint centred at 67.31$^\circ$N, 348.92$^\circ$E). MASCS data at wavelengths exceeding 800 nm are not used in the present study.}
   \label{fig1}
   \end{figure}

Requiring that data are available at each of the eight retained filter wavelengths leaves 16,064 images (8$\times$2008 images of a same area) for constructing an 8-band map of Borealis using ISIS. Image processing implies the following steps (\citealt{bott2018shakespeare}): to import raw data into ISIS format and to convert them to reflectance; to perform radiometric calibration to correct for bias, dark current and flat field; to apply polar stereographic projection with 450 m/pixel resolution; to apply Kaasalainen-Shkuratov corrections using so-called “standard” reference values, $i_{ref}=30$$^\circ$, $\varphi_{ref}=30$$^\circ$, and $e_{ref}=0$$^\circ$; to co-register each image with the 749$\;\mathrm{nm}$ filter; to stack and trim images in order to obtain the mosaic map of Borealis. Pixels having reflectance lower than 0.005 or larger than 0.3 have been discarded in the following analysis. As an example, Figure \ref{fig1} (left) displays the map of Borealis obtained with the 996 nm filter. The distributions of the reflectance measured at each of the selected wavelengths are fitted to a Gaussian with mean and standard deviation values listed in Table \ref{table1}. Two of these, at 433 nm and 996 nm, are displayed in Figure \ref{fig2} as examples.

%-------------------------------------------------------------------------
\subsection{MASCS Data}
\label{section2.2}
MASCS includes two spectrometers: the UltraViolet and Visible Spectrometer (UVVS, 115-600 nm) and the Visible and InfraRed Spectrometer (VIRS, 300-1450 nm). UVVS determines the composition and structure of Mercury’s exosphere and measures the surface reflectance, while VIRS measures only the surface reflectance. The present analysis uses only observations made by the latter. VIRS is a point spectrometer with a 0.023$^\circ$ impling a footprint covering typically a few 100 metres. The data of the near-infrared detector (900-1450 nm) are known to be noisy (\citealt{besse2015spectroscopic}) and are not used in the present work, which retains only data of the visible detector (300-1050 nm) at wavelengths smaller than 800 nm. Photometric corrections do not use standard reference values but use instead $\varphi_{ref}=90^\circ$, $e_{ref}=45^\circ$, and $i_{ref}=45^\circ$ (\citealt{izenberg2014low, besse2015spectroscopic}). A typical spectrum is shown in Figure \ref{fig1} (right) as example.

\begin{figure}
   \centering
   \includegraphics[width=14cm]{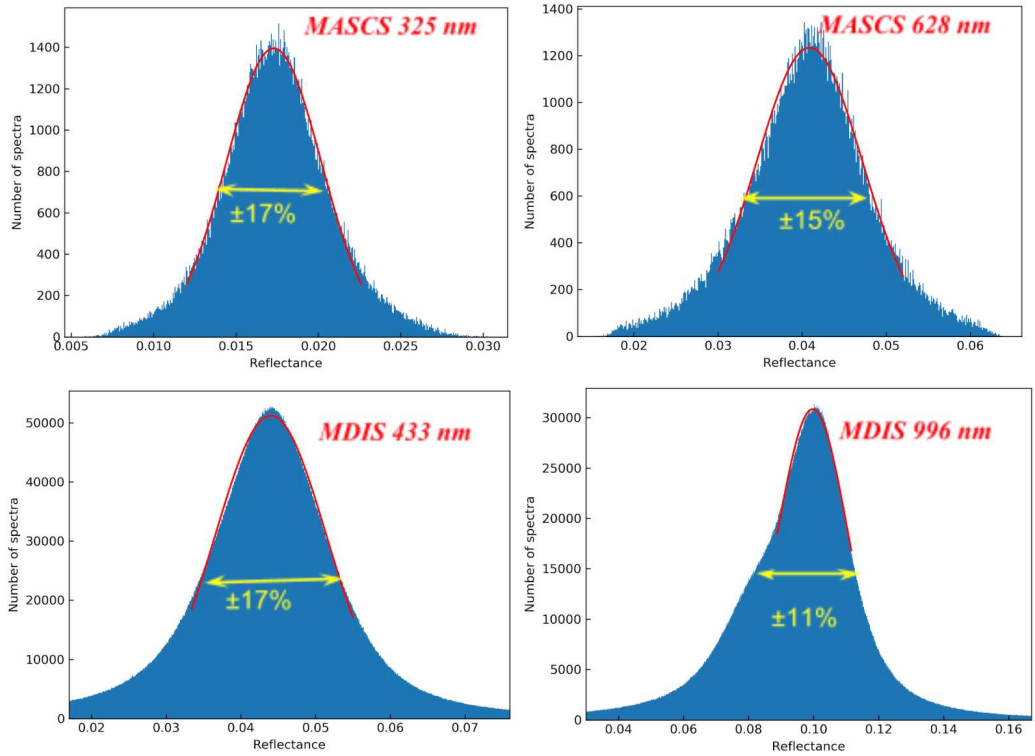}
   \caption[]{Examples of reflectance distributions of MASCS (upper panels) and MDIS (lower panels) data with Gaussian fits (red curves, Table \ref{table1}). The wavelengths, selected in the lower and upper parts of the wavelength range respectively, are indicated in the inserts. Yellow arrows are at half-height and labelled with the value of the $\sigma$/mean ratio.}
   \label{fig2}
   \end{figure}

The MASCS data are in the form of spectra using an average bin width of 4 nm. For comparison, the filters used in the MDIS analysis cover between 5 and 18 nm. In the present work, we select nine wavelengths for subsequent analysis: four wavelengths (300, 310, 325 and 390 nm) matching those normally used to define spectral parameters in the UV range and five wavelengths (432, 479, 553, 628, and 748 nm) matching the MDIS filters in the interval of wavelength where data overlap. As for MDIS data, the distributions of the reflectance measured at each of the selected wavelengths are fitted to a Gaussian with mean and standard deviation values listed in Table \ref{table1}. Figure \ref{fig2} shows, as examples, the reflectance distributions at 325 nm and 628 nm overlaid with the Gaussian fits.

\begin{table}
\begin{center}
\caption[]{MASCS and MDIS data: dependence on wavelength of the mean and $\sigma$ values of the measured reflectance (in per mil).}\label{table1}
 \begin{tabular}{c c c c c c c c c c}
 \hline
\multicolumn{10}{c}{\textbf{MASCS data}}\\ \hline
$\lambda$ (nm)&300&310&325&390&432&479&553&628&748\\ \hline
Mean ($10^{-3}$)&15.5&14.6&17.3&23.8&27.8&29.6&35.5&40.9&48.4\\ \hline
$\sigma$ ($10^{-3}$)&2.7&2.4&2.9&4.0&4.6&4.8&5.6&6.3&7.4\\ \hline
\end{tabular}
\begin{tabular}{c c c c c c c c c}
\multicolumn{9}{c}{\textbf{MDIS data}}\\ \hline
$\lambda$ (nm) &433&480&559&629&749&828&899&996\\ \hline
Mean ($10^{-3}$)&44.0&52.1&61.4&70.3&81.0&93.0&97.8&99.7\\ \hline
$\sigma$ ($10^{-3}$)&7.5&8.3&9.4&9.7&10.2&10.6&11.0&10.9\\ \hline
\end{tabular}
\end{center}
\end{table}

%-------------------------------------------------------------------------
\subsection{Comparing MDIS with MASCS Data}
\label{section2.3}
The results listed in Table \ref{table1} call for a number of important comments. 
 
In the range of wavelengths where both MASCS and MDIS reflectance measurements are available, the ratio between the latter and the former is 1.69$\pm$0.06. We remark that the ratio between MDIS and MASCS reflectance is nearly independent of wavelength, fluctuating by less than 4\% (0.06/1.69) over the 316 nm explored range. This suggests that the difference is simply the result of a global rescaling, a same ratio at all wavelengths. 
	
	We note that the MDIS/MASCS ratio measured in the Caloris Basin (\citealt{besse2015spectroscopic}) is $\sim$1.8 for hollows of Tyagaraja and $\sim$1.3 for Eminescu crater, suggesting a dependence on the nature of the terrain being probed. However, in the present study, we shall ignore such dependence as it has only minor influence on our results and conclusions.
	
	The ratio between the standard deviation ($\sigma$) and mean values of the distributions of measured reflectance is also remarkably constant over the respective explored ranges of wavelength, 16.3$\pm$0.7\% for MASCS data and 13.5$\pm$2.2\% for MDIS data. Over the range of overlapping wavelengths, these numbers become 15.8$\pm$0.5\% and 14.9$\pm$1.6\% respectively. Their similarity suggests that the measured values of the reflectance are only weakly affected by measurement uncertainties, most of their variation between different measurements being of physical origin.
	  
	The above ratios, at the level of $\sim$15\% on average, are largely due to the spread of average reflectance values, averaged over wavelengths, rather than to a difference of shape, relative slope or curvature. Indeed, when defining a normalised reflectance as the ratio between the measured reflectance and the mean reflectance averaged over all wavelengths, the above ratios, evaluated on the normalised reflectance distributions, become much smaller, typically at the 5\% level. As a result, we expect a strong correlation between the mean reflectance and the mean spectral slope of a given measurement. In what follows we distinguish between the absolute spectral slope, proportional to the derivative of the reflectance with respect to wavelength, and the relative spectral slope, proportional to its logarithmic derivative. 
 	
	As can be seen from Table \ref{table1}, the mean reflectance of the surface of Borealis is low, between 1.5\% and 5\% for MASCS data, between 4.4\% and 10\% for MDIS data. Both mean and standard deviation values increase with wavelength, faster at lower than at higher wavelengths, the mean values by some 80\% over the $\sim$320 nm interval of overlapping wavelengths.

%-------------------------------------------------------------------------
%-------------------------------------------------------------------------
\section{MDIS Data}

The number of selected wavelengths, eight for MDIS and nine for MASCS, is larger than the number of independent quantities that can be expected to carry relevant information. In order to ease the interpretation of the reflectance data and to correlate their spectral variations with compositional heterogeneities of the Mercury’s surface, it is useful to define a small number of parameters characterizing the spectral dependence. One way to do so is simply to select a few wavelengths, as done in false colour red-green-blue (RGB) maps. A better method, making in principle optimal use of the available information, is the Principal Component Analysis method (PCA) (e.g., \citealt{richards1999remote}). More generally, we need to find how many parameters need to be used in order to characterize the measurements reliably. Several different parameters have been used in earlier studies, but many of these are not independent from each other. In the present section, we concentrate on MDIS data; in a first phase, we review their interpretations in terms of spectral parameters used in earlier studies. Next, we give a unified picture of these different approaches and illustrate its relation to the geological and morphological properties of the regolith.

\subsection{Descriptions in Terms of Spectral Parameters Used in Earlier Studies}
\label{section3.1}
\begin{figure}
  \centering
  \includegraphics[width=15cm]{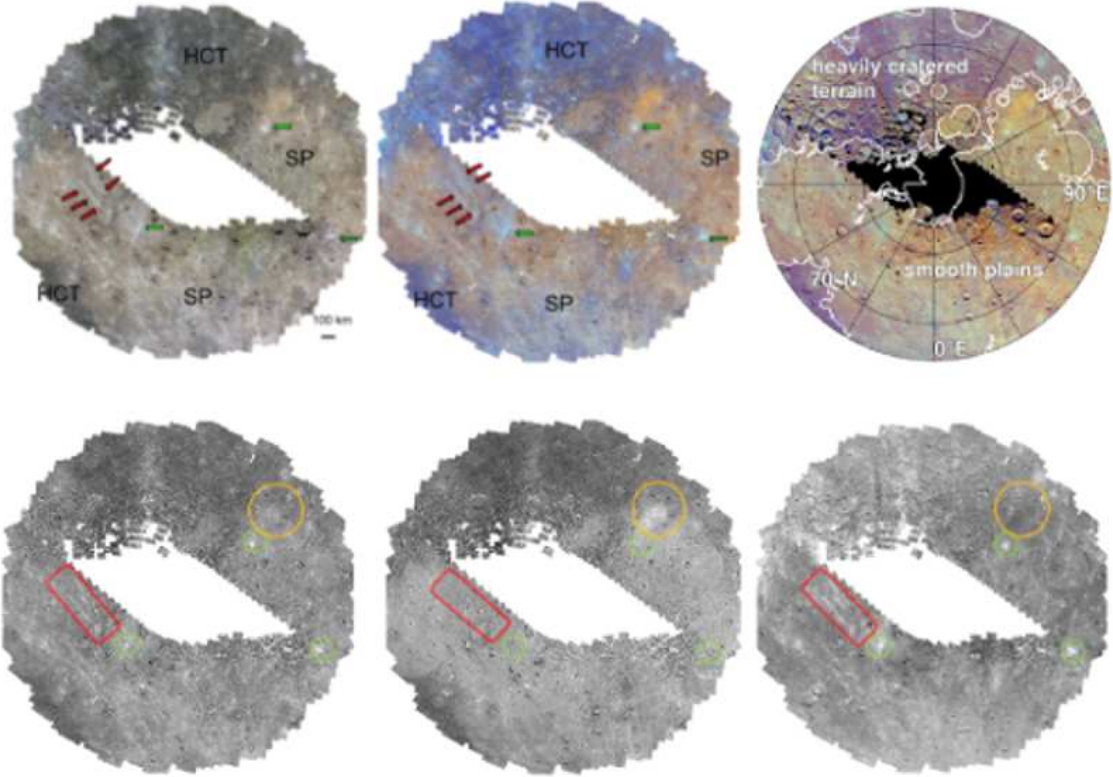}
  \caption{MDIS maps of Borealis. Upper-left: False colour RGB map (R$=$996 nm, G$=$749 nm and B$=$433 nm). Upper-central: Enhanced colour map (Red$=$\textit{PC2}, Green$=$\textit{PC1}, Blue$=$$\rho$). Upper-right: Enhanced colour map from \citealt{ostrach2017geologic}. Red arrows point to crater rays and green arrows to fresh craters. Lower panels (from left to right): separate maps of \textit{PC1}, \textit{PC2} and $\rho$. Fresh ejecta (green), crater rays (red) and RPG (orange) are delineated.}
  \label{fig3}
  \end{figure}

Figure \ref{fig3} (upper left) displays a false colour RGB map of Borealis using three filters R$=$996 nm, G$=$749 nm and B$=$433 nm, the darker areas showing the northern HCTs with high density of craters and the brighter areas showing the northern SPs. The enhanced colour map obtained from PCA is displayed in the upper central panel and is compared with that obtained by \citealt{ostrach2017geologic} (upper right). More precisely the latter two maps use the first principal components \textit{PC1} and \textit{PC2} as green and red, respectively, but rather than using \textit{PC3} as blue, they use instead the reflectance ratio $\rho$ between the 433 nm and 996 nm values as was done previously by other authors. The appearance obtained with the enhanced colour map allows for a clearer distinction between different types of terrains than the false colour RGB map does, although both maps carry essentially the same information. In particular, the distinction between SP and HCT areas becomes clearer. The orange areas identify the SPs, the dark blue areas identify the HCTs, brightness being an indicator of fresher materials. Bright blue is associated with materials from young impact craters such as crater rays (red arrows) from Hokusai (\citealt{xiao2016self}) and fresh ejecta around fresh craters (green arrows). 

\begin{figure}
  \centering
  \includegraphics[width=15cm]{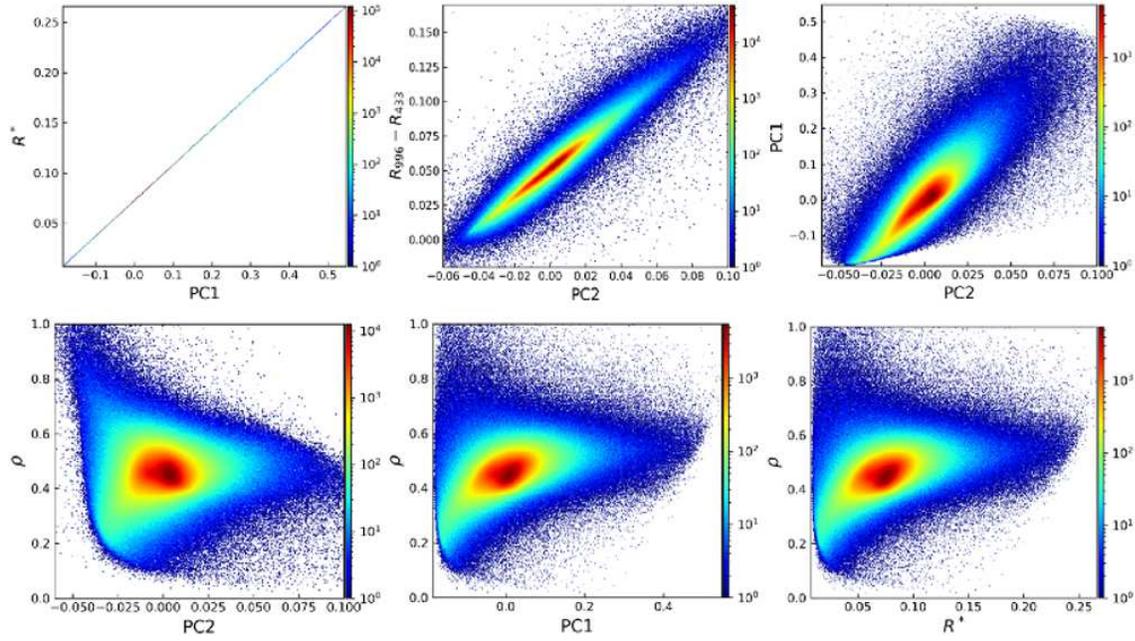}
  \caption{Correlation between different spectral parameters (note the logarithmic colour scales): from left to right and up down, $R^*$ vs \textit{PC1}, $R_{996}-R_{433}$ vs \textit{PC2}, \textit{PC1} vs \textit{PC2}, $\rho$ vs \textit{PC2}, $\rho$ vs \textit{PC1} and $\rho$ vs $R^*$.}
  \label{fig4}
  \end{figure}

\textit{PC1} is known to highlight brightness and \textit{PC2} to highlight the absolute spectral slope. This is clearly illustrated in Figure \ref{fig4} that displays the correlation between different spectral parameters; \textit{PC1} is approximately equal to $0.3\,(R^*/\langle R^* \rangle -1)$ where $R^*$ is the brightness, defined for each measured spectrum as the reflectance averaged over all wavelengths, and where $\langle R^* \rangle$ is its mean, averaged over all measurements, $\langle R^* \rangle \sim0.073$; \textit{PC2} is nearly equal to $-0.05+R_{996}-R_{433}$, namely depends only on the absolute spectral slope: \textit{PC1} and \textit{PC2} are therefore trivially correlated. On the contrary, 
$\rho=R_{433}/R_{996}=[(R_{996}-R_{433})/R_{433}+1]^{-1}$, 
depends only on the relative spectral slope, $(R_{996}-R_{433})/[(996-433)R_{433}]$ and displays very little correlation with \textit{PC1}, \textit{PC2} and $R^*$. 

Separate maps of \textit{PC1}, \textit{PC2} and $\rho$ are displayed in the lower panels of Figure \ref{fig3}. Inspecting them separately shows that most features are visible on each of them but are differently enhanced. Globally, the \textit{PC1} and \textit{PC2} maps distinguish clearly between SPs and HCTs, at strong variance with the $\rho$ map. However, fresh ejecta around fresh craters and crater rays are almost invisible in the \textit{PC2} map but are seen to display both a high brightness (\textit{PC1}) and a low relative spectral slope (high $\rho$ value). On the contrary, the RPG is particularly enhanced in the \textit{PC2} map and barely visible on the $\rho$ map. The latter observations may seem to contradict the former: globally, $\rho$ stands out as carrying information independent from \textit{PC1} and \textit{PC2} but, when looking at some specific features, \textit{PC2} stands out as carrying information independent from \textit{PC1} and $\rho$.  As $\rho$ varies approximately as $PC1/PC2$, and therefore \textit{PC2} approximately as $PC1/\rho$, the RPG corresponds to high \textit{PC1} and small $\rho$, and is therefore enhanced on the \textit{PC2} map; the crater rays and fresh ejecta correspond to high \textit{PC1} and high $\rho$, but their \textit{PC2} values are average. The SPs correspond to high \textit{PC1} and high \textit{PC2}, the HCTs to low \textit{PC1} and low \textit{PC2}, but their $\rho$ values are average.

\begin{figure}
  \centering
  \includegraphics[width=15cm]{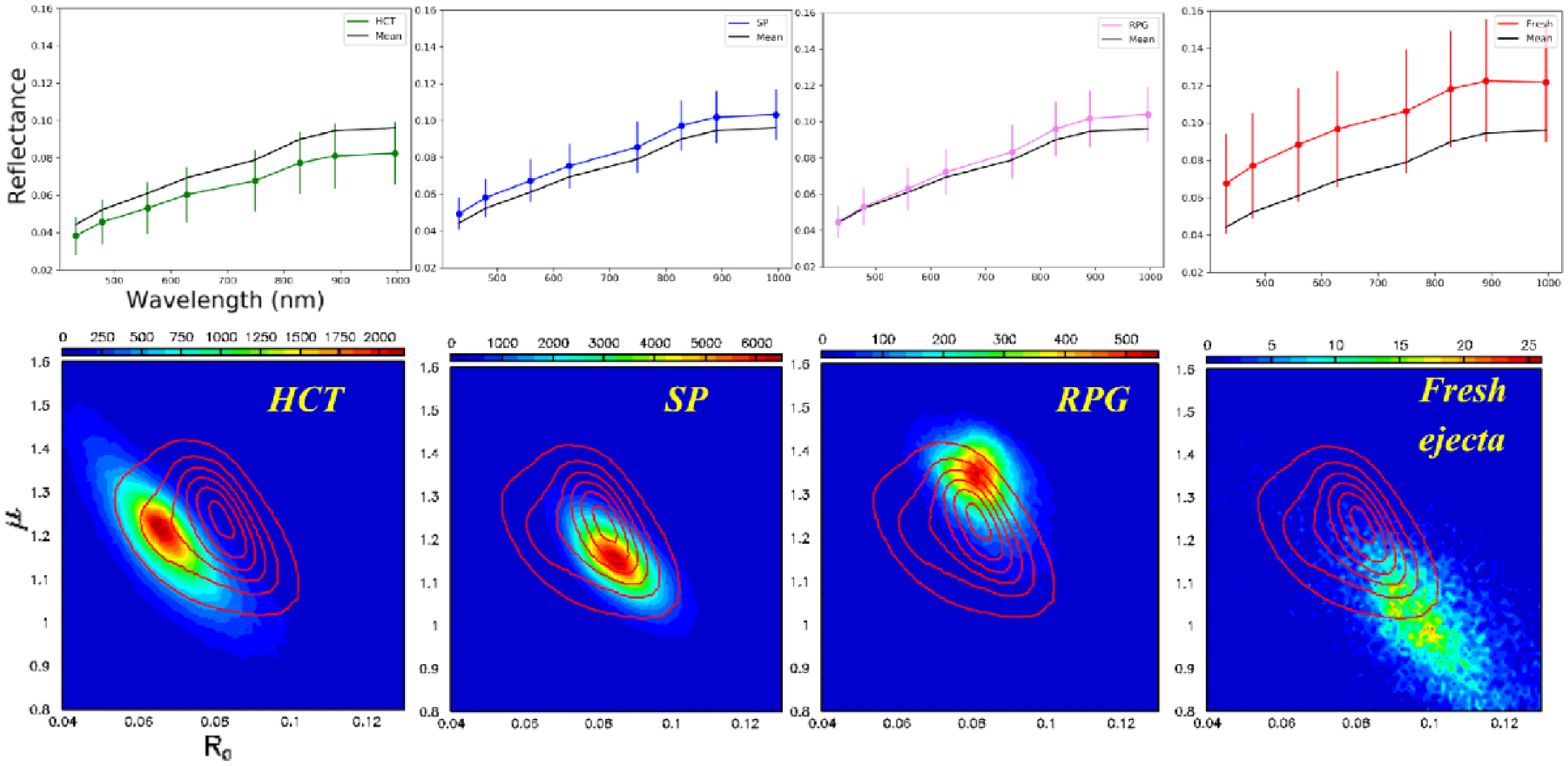}
  \caption{MDIS spectra associated with the regions listed in Table \ref{table2}. Upper panels: mean spectra (HCT$=$green, SP$=$blue,  RPG$=$magenta, Fresh$=$red); error bars indicate the standard deviation and the mean spectrum associated with the whole Borealis region is shown in black. Lower panels: regions of the $\mu$ vs $R_0$ plane populated by the selected geological regions (see Section \ref{section3.2}). The contours correspond to the whole Borealis region (in steps of 1 from 1 to 6 in relative units).}
  \label{fig5}
  \end{figure}

In order to clarify this point, we select four specific regions on the Borealis map and display in Figure \ref{fig5}, for each of these, the dependence of the reflectance on wavelength. The selected regions are defined by an interval of latitude and an interval of longitude as listed in Table \ref{table2}. Each of the obtained spectra is fitted to a second degree polynomial of the form $R=R_0[1+\mu(\lambda-\lambda_0)+\nu(\lambda-\lambda_0)^2]$ with $\lambda_0=715$ nm in the middle of the explored wavelength interval. The values of $R_0$, $\mu$ and $\nu$ are listed in Table \ref{table2} together with the values obtained for the whole map. Here, $R_0$ measures the brightness, $\mu$ the relative spectral slope and $\nu$ the curvature of the spectrum. As can be verified on Table \ref{table2}, $R_0$ corresponds to \textit{PC1} and $\mu$ to $1/\rho$ while $\nu$ varies by only $\pm$5\% about its mean value of $-1.22$.

\begin{table}
\begin{center}
\caption[]{Selected representative regions. Longitude intervals are defined spanning clockwise from the lower to higher limits. Fresh ejecta are from craters located at $\sim$80$^\circ$N, 47.5$^\circ$W and 68.5$^\circ$N, –69.5$^\circ$E, respectively. Second degree polynomial parameters $R_0$, $\mu$, and $\nu$ representing the brightness, relative spectral slope, and spectral curvature, respectively, are listed in the upper row for MDIS and in the lower row for MASCS (with $\lambda_0=715$ nm for MDIS data and 600 nm for MASCS data).}\label{table2}
\begin{tabular}{|c|c|c|c|c|c|c|c|}

\hline
\multirow{2}{*}{  }&\multicolumn{2}{c|}{longitude}&\multicolumn{2}{c|}{latitude}&$R_0$&$\mu$&$\nu$\\
\cline{2-3}\cline{4-5}&from&to&from&to&($10^{-3}$)&($10^{-3}$)&($10^{-6}$)\\

\hline
\multirow{2}{*}{HCT}&\multirow{2}{*}{180$^\circ$}&\multirow{2}{*}{150$^\circ$E}&\multirow{2}{*}{65$^\circ$N}&\multirow{2}{*}{75$^\circ$N}&67.9 &1.20&–1.16 \\
&&&&&33.8& 1.56&–1.51\\

\hline
\multirow{2}{*}{SP}&\multirow{2}{*}{30$^\circ$E}&\multirow{2}{*}{30$^\circ$W}&\multirow{2}{*}{65$^\circ$N}&\multirow{2}{*}{75$^\circ$N}&83.3 &1.16&–1.15 \\
&&&&&43.2& 1.55&–1.42\\

\hline
\multirow{2}{*}{RPG}&\multirow{2}{*}{140$^\circ$E}&\multirow{2}{*}{125$^\circ$E}&\multirow{2}{*}{72$^\circ$N}&\multirow{2}{*}{77$^\circ$N}&82.8 &1.32&–1.10 \\
&&&&&42.9& 1.81&–1.20\\

\hline
\multirow{2}{*}{Fresh}&\multicolumn{4}{c|}{\multirow{2}{*}{see caption}}&106.8 &0.97&–1.27 \\
&\multicolumn{4}{c|}{}&53.1&1.39&–1.80\\

\hline
\multirow{2}{*}{All}&\multicolumn{2}{c|}{\multirow{2}{*}{all}}&\multirow{2}{*}{65$^\circ$N}&\multirow{2}{*}{80$^\circ$N}&78.5 &1.22&–1.14 \\
&\multicolumn{2}{c|}{}&&&38.4& 1.69&–1.27\\
\hline
\end{tabular}
\end{center}
\end{table}

%-------------------------------------------------------------------------
\subsection{Unified Picture}
\label{section3.2}
\begin{figure}
  \centering
  \includegraphics[width=14cm]{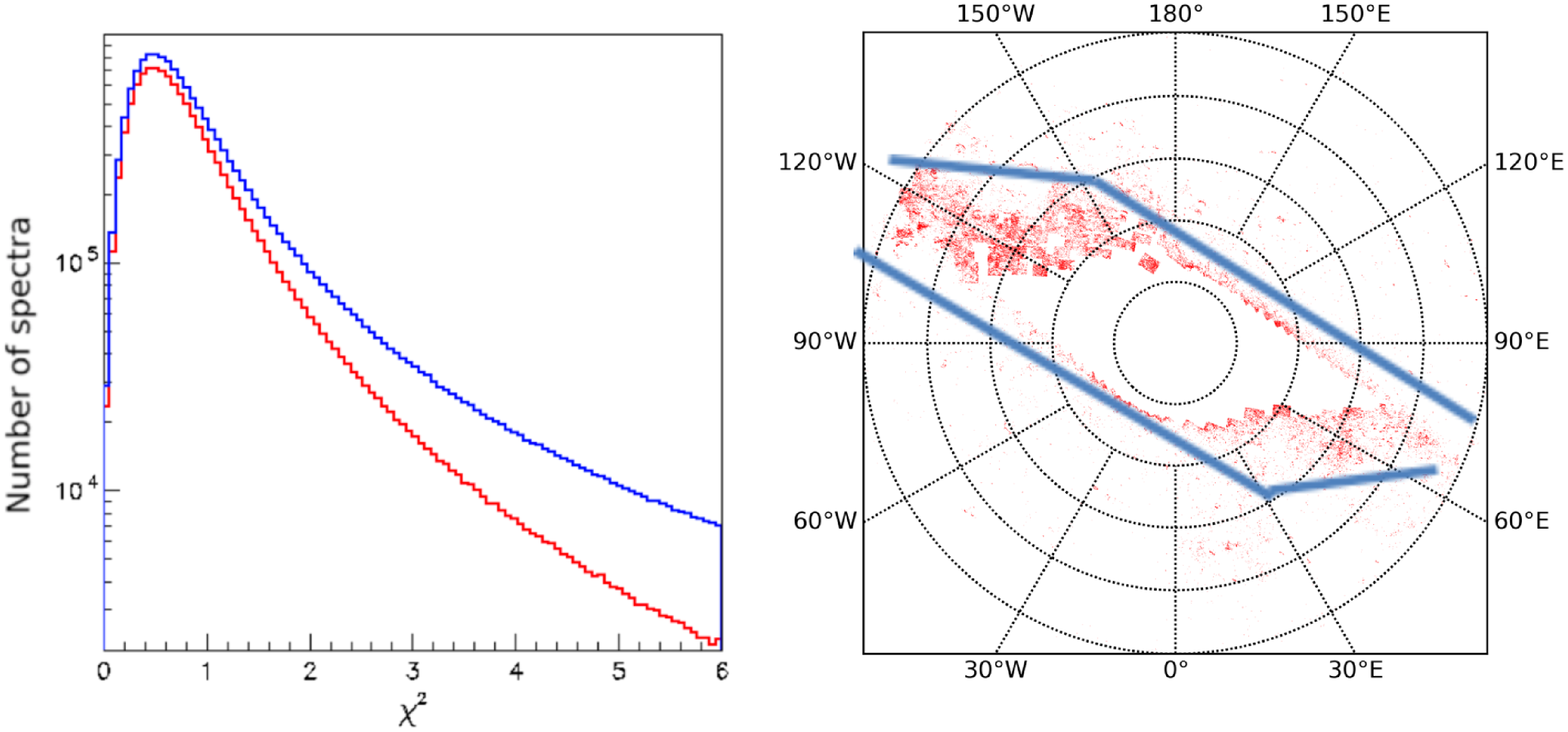}
  \caption{Second degree polynomial fits to the MDIS reflectance spectra. Left: $\chi^2$ distribution including (blue) and excluding (red) the polar zone delineated in the right panel. Right: location on the Borealis map of spectra giving a $\chi^2$ in excess of 6.}
  \label{fig6}
  \end{figure}

The above results suggest extending to all measured spectra the second degree polynomial description used for the average spectra of selected regions (Table \ref{table2}). In doing so, we calculate for each spectrum the $\chi^2$ (normalised to the number of degrees of freedom) of the fit using an uncertainty of $3\times10^{-3}$ on each reflectance measurement. Its distribution is shown in blue in the left panel of Figure \ref{fig6}. It displays a long tail associated with polar regions where the incidence angle is largest, as shown in the right panel of the figure. For the sake of the arguments developed in the present section, it is more important to deal with a clean data sample than to cover as much as possible of the Borealis surface. Consequently we exclude from the following analysis the region delineated by the dark blue line in the right panel of Figure \ref{fig6}. The resulting distribution of $\chi^2$ is shown in red in the left panel; its mean value has decreased from 1.07 to 0.92 and its rms value from 0.95 to 0.77.

\begin{figure}
  \centering
  \includegraphics[width=14cm]{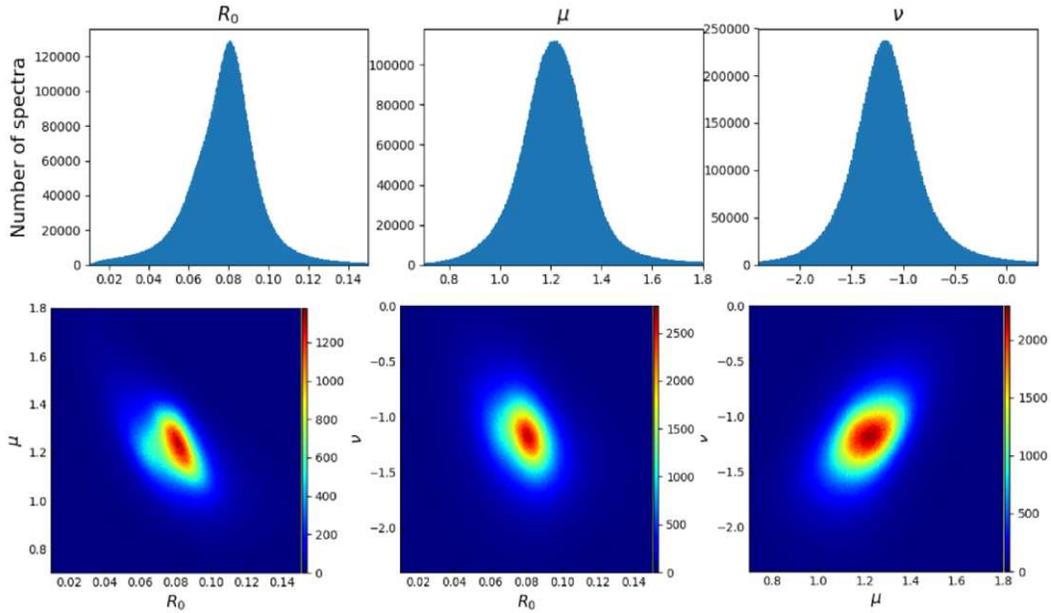}
  \caption{Second degree polynomial parameters fitting the MDIS reflectance spectra. Upper panels: distributions of brightness $R_0$ (left), relative spectral slope $\mu$ (central) and spectral curvature $\nu$ (right). Lower panels: correlation between $\mu$ and $R_0$ (left), $\nu$ and $R_0$ (central), and $\nu$ and $\mu$ (right).}
  \label{fig7}
  \end{figure}

The resulting distributions and correlations of the three parameters, $R_0$, $\mu$ and $\nu$, are displayed in Figure \ref{fig7}. The population of the $\mu$ vs $R_0$ plane associated with each of the selected regions listed in Table \ref{table2} is illustrated in the lower panels of Figure \ref{fig5}.

The obtained characterization of the measured reflectance spectra confirms what had been revealed by the considerations developed earlier. Different spectra differ mostly by the value of the mean reflectance, measured by $R_0$. In particular, such is the case for the two main types of terrains, SPs and HCTs; the values taken by $\mu$ and $\nu$, as can be seen from Table \ref{table2} and Figure \ref{fig5}, do not help much with distinguishing between them. However, once a class of terrain has been defined on the basis of not only the spectral parameters but also other considerations of a geological and/or morphological nature, it is seen to occupy a specific location in the spectral parameter space. Indeed, the shape of the region occupied in this space by the totality of measured spectra reveals clearly the existence of two overlapping classes, one corresponding to SPs and the other to HCTs. But in the overlap region it is not possible, on the sole basis of the values taken by the spectral parameters, to decide to which class a given spectrum belongs. A definition of SPs and HCTs in terms of the value taken by $R_0$, or more generally in exclusive terms of the spectral dependence of the surface reflectance, can only be arbitrary; other considerations, such as the density of craters, need to be taken in consideration. 

This result is very general; the reflectance spectra are too similar, in particular once normalised to a same $R_0$ value, to allow for any meaningful separation into different classes; the error bars displayed in the upper panels of Figure \ref{fig5} are too large in comparison with the separation between the different spectra; the areas covered in the $\mu$ vs $R_0$ plane displayed in the lower panels of Figure \ref{fig5} are too large with respect to the separation between different geological regions. Fresh ejecta, for example, that are clearly identified from their high $R_0$ value, are in the continuity of the SP population and overlap it: one cannot define a value of $R_0$ that would reliably distinguish them from bright SPs; such a distinction requires additional arguments based on other criteria than the spectral dependence of the reflectance. The same is true of the RPG, in spite of its larger average value of the relative spectral slope $\mu$; its population overlaps that of SPs, with no clear separation between the two types. This result is also confirmed by an attempt that we made to define different terrain classes using the so-called K-means method (\citealt{macqueen1967some}); all what can be obtained is essentially an arbitrary separation between slices covering different intervals of $R_0$.   

\begin{figure}
  \centering
  \includegraphics[width=14cm]{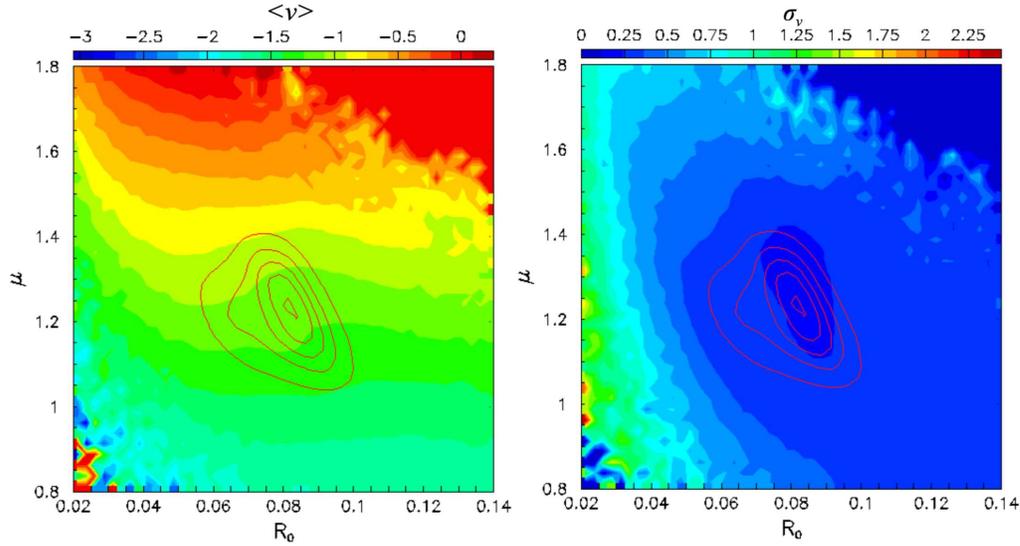}
  \caption{Curvature of the second degree polynomial fits to the MDIS reflectance spectra. Distributions in the $\mu$ vs $R_0$ plane of the mean value of $\nu$, $\langle \nu \rangle$, and of its standard deviation $\sigma_{\nu}$, are shown in the left and right panels respectively. Contours show the population of reflectance spectra.}
  \label{fig8}
  \end{figure}

In this sense, we can state that two parameters, $R_0$ and $\mu$ are sufficient to completely characterize the spectral dependence of the reflectance in the Borealis quadrangle. As additional evidence that the curvature parameter $\nu$ does not carry any additional information once the values of $R_0$ and $\mu$ are known, we display in Figure \ref{fig8} the dependence on $R_0$ and $\mu$ of its mean value $\langle \nu \rangle$ and standard deviation $\sigma_{\nu}$. It shows that $\nu$ is essentially independent of $R_0$ but increases with $\mu$ as was already apparent from the correlation plots in Figure \ref{fig7}. Its dispersion is observed to be very small, typically 15\% of its absolute value, adding confidence to our conclusions. 

  \begin{figure}
  \centering
  \includegraphics[width=13cm]{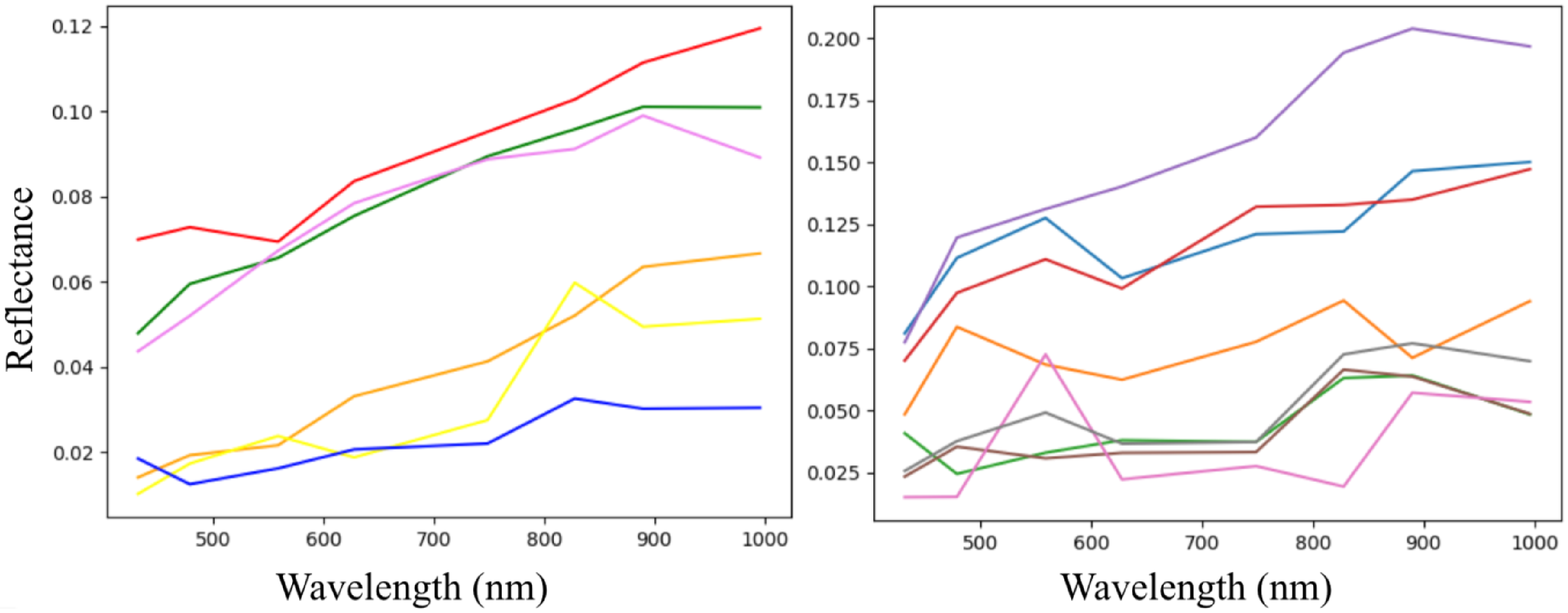}
  \caption{MDIS reflectance spectra having $\chi^2>6$. Left: a random sample. Right: selected large $\chi^2$ values around 120$^\circ$W longitude.}
  \label{fig9}
  \end{figure}

%-------------------------------------------------------------------------
\subsection{Deviation from a Simple Polynomial Description of the Spectral Dependence of the Reflectance}
\label{section3.3}
In the preceding section, we excluded from the analysis a region of the Borealis surface hosting many spectra having a large value of $\chi^2$, namely deviating significantly from a second degree polynomial fit. Such a deviation may be associated with a real feature, such as absorption in a specific wavelength region, or simply result from the smearing caused by measurement errors. In the present section, we address this issue. By inspecting examples of large $\chi^2$ spectra, we observe (Figure \ref{fig9}) that some of these, located around 120$^\circ$W longitude, seem to display a depression around 0.7 $\mu$m.

\begin{figure}
  \centering
  \includegraphics[width=15cm]{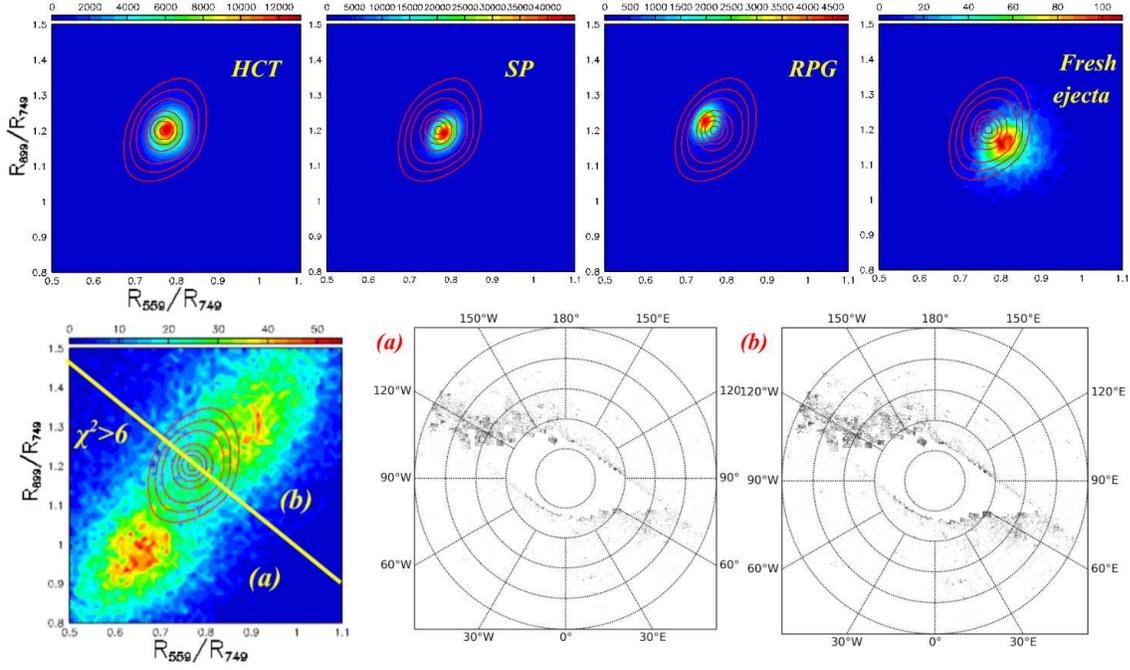}
  \caption{Upper and lower left panels: distributions of different spectra in the $R_{high}$ vs $R_{low}$ plane as indicated in the inserts. Contours show the distribution over the whole Borealis region (at 1, 2, 4, 12, 20, 28, 36 in relative units). The lower right panels show the maps of the pixels with spectra having $\chi^2>6$ for regions of the $R_{high}$ vs $R_{low}$ plane marked as (a) and (b) in the lower left panel.}
  \label{fig10}
  \end{figure}

We define accordingly two ratios, $R_{low}=R_{559}/R_{749}$ and $R_{high}=R_{899}/R_{749}$. On average, for good $\chi^2$ spectra, we expect them to be independent of $R_0$, their product to be independent of $\mu$, and their ratio to be a measure of $\mu$. This is indeed what we observe when displaying, in Figure \ref{fig10}, the distribution of different spectra in the $R_{high}$ vs $R_{low}$ plane. The global distribution and that associated with the selected regions listed in Table \ref{table2} are all confined around $R_{low}\sim0.78$ and $R_{high}\sim1.20$ as expected, with a small spread corresponding to their different values of $\mu$. On the contrary, large $\chi^2$ spectra cover a very broad region where $R_{high}\sim1.2+1.3(R_{low}-0.8)\pm0.1$. This corresponds to spectra where the 749 nm reflectance is either larger than expected from the polynomial fit (labelled (a) on the figure) or smaller than expected from the polynomial fit (labelled (b) on the figure). The central depression corresponds to spectra for which the 749 nm reflectance matches the polynomial fit, associated with a low $\chi^2$ value. The maps of pixels having spectra in the regions (a) and (b) are shown separately in the lower right panels of Figure \ref{fig10}. They are very similar, implying that the cause of the large $\chi^2$ values is purely instrumental and does not reveal any special feature of morphological or geological relevance.

Table \ref{table3} lists the mean and standard deviation values of the distributions of the difference between the measured and fitted values of the reflectance at each of the eight wavelengths where measurements are available. The mean values show significant deviations, $\pm1.3\times10^{-3}$ on average, that correspond to modulations, with a maximum of $2.7\times10^{-3}$ at 749 nm. One could of course refine the fit by including these modulations in the model, but this would not contribute significant information to the questions being addressed here. The standard deviations are of similar magnitudes, $1.2\times10^{-3}$ on average, again with a maximum of $1.5\times10^{-3}$ at 749 nm. These results illustrate the quality of the fits and provide additional evidence for the absence of a special feature.

\begin{table}
\begin{center}
\caption[]{Distributions of the difference between the measured and fitted MDIS reflectance.}\label{table3}
\begin{tabular}{c c c c c c c c c c}
 \hline
$\lambda$ (nm)&433&480&559&629&749&828&899&996\\ 
 \hline
 Mean ($10^{-3}$)&$-$0.2&1.1&$-$0.6&$-$0.4&$-$2.7&2.0&2.3&$-$1.5\\			
$\sigma$ ($10^{-3}$)&0.9&1.0&1.3&1.4&1.5&1.4&1.4&0.8\\
 \hline
\end{tabular}
\end{center}
\end{table}

%-------------------------------------------------------------------------
%-------------------------------------------------------------------------
\section{MASCS Data}
\subsection{Descriptions in Terms of Spectral Parameters Used in Earlier Studies}
\label{section4.1}
As was done for MDIS data, we start with following the method previously developed by other authors, using four spectral parameters, two in the visible range, and two at shorter wavelengths in the UV range. However, we find that the latter are too noisy to allow for a clear correlation with geological features and we present only the results obtained in the visible: we retain as spectral parameters \textit{VIS ratio}, defined as the reflectance ratio between the 410 nm and 750 nm measurements and \textit{VIS slope}, defined as the reflectance difference between the 750 nm and 445 nm measurements, divided by the associated wavelength interval of 305 nm. \textit{VIS ratio} is commonly used to compare observations with mean planetary MASCS spectra (\citealt{izenberg2014low}) and \textit{VIS slope} is used to study spectral characteristics of pyroclastic deposits on Mercury (\citealt{besse2015spectroscopic}).

\textit{VIS slope}, $6.51\times10^{-5}$ nm$^{-1}$ on average, is measured in units of the mean \textit{VIS slope}, averaged over all MASCS measurements obtained on Mercury, $6.27\times10^{-5}$ nm$^{-1}$. It measures the dependence of the reflectance on wavelength in absolute rather than relative terms: as was discussed in Section \ref{section2.3}, it is trivially positively correlated to the mean reflectance, or, for that matter, to the value of the reflectance at any representative wavelength, such as, for example, 553 nm, a reference commonly used by other authors. \textit{VIS ratio}, 0.53 on average, measures instead the dependence of the reflectance on wavelength in relative terms and is therefore expected to display relatively smaller variations than \textit{VIS slope} and not to be trivially correlated to $R_{553}$ as \textit{VIS slope} is. However, if \textit{VIS slope} is larger or smaller than what is expected from simple rescaling in proportion to $R_{553}$, one expects \textit{VIS ratio} to be respectively smaller or larger than average: $R_{553}$ and \textit{VIS ratio} are independent variables but \textit{VIS slope} is correlated with both.     

Spectra having values of $R_{553}$ and of the two VIS spectral parameters too far from average are removed from further analysis. More precisely, the accepted intervals are 0.015 to 0.055 for $R_{553}$, 0.48 to 0.65 for \textit{VIS ratio} and 0.4 to 1.7 for \textit{VIS slope}. There remain 189,860 spectra after application of the cuts.  

\begin{figure}
  \centering
  \includegraphics[width=15cm]{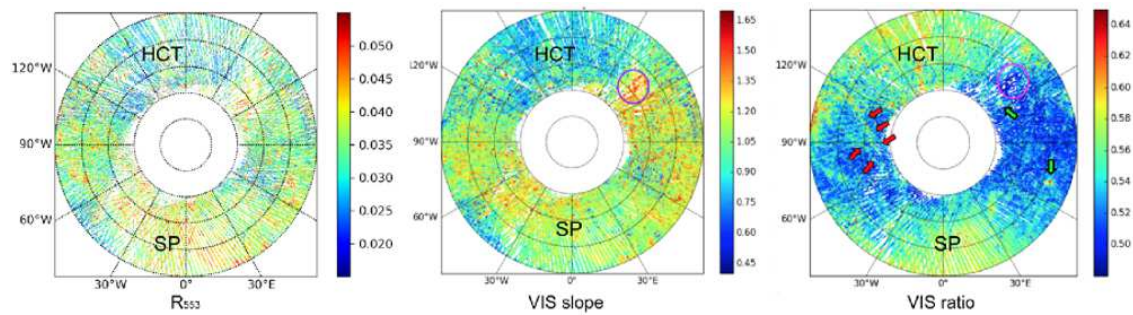}
  \caption{MASCS maps of Borealis: $R_{553}$ (left), \textit{VIS slope} (middle) and \textit{VIS ratio} (right). Circles show the RPG region.}
  \label{fig11}
  \end{figure}

Figure \ref{fig11} maps each of the three spectral parameters, $R_{553}$, \textit{VIS slope} and \textit{VIS ratio}, separately. They distinguish clearly between different areas. Each measurement is plotted on the map at the position of the centre of the footprint. Blank  correspond to missing data. \textit{VIS slope} distinguishes between the two major geological units, HCTs, having smaller values (shown in blue) and SPs having higher values (shown in yellow and red). The larger values are reached in the bright red RPG region indicated by a circle on the map. HCTs and part of SPs display high values of \textit{VIS ratio} while several Hokusai crater rays (red arrows) are visible in the low \textit{VIS ratio} regions. The SPs are seen to be split into two low \textit{VIS ratio} regions bracketing a high \textit{VIS ratio} region. In very fresh craters (green arrows), \textit{VIS ratio} is higher at the centre and lower on the periphery.

MASCS and MDIS data cover overlapping but different ranges of wavelengths; however, at least qualitatively, we expect the triplet \textit{\{$R_{553}$, VIS slope, VIS ratio\}} in MASCS data to carry similar information as the triplet $\{PC1,PC2,1/\rho\}$ does in MDIS data. Indeed, the RPG, enhanced on the \textit{PC2} map  is also enhanced on the \textit{VIS slope} map; crater rays and fresh ejecta, with high \textit{PC1}, low 1/$\rho$ and average \textit{PC2} have high $R_{553}$, low \textit{VIS ratio} and average \textit{VIS slope}; SPs with high \textit{PC1}, high \textit{PC2} and average 1/$\rho$ have high $R_{553}$, high \textit{VIS slope} and average $\textit{VIS ratio}$; HCTs, with low \textit{PC1}, low \textit{PC2} and average $\rho$ have low $R_{553}$, low \textit{VIS slope} and average \textit{VIS ratio}. However, the strong east and west depressions displayed by the \textit{VIS slope} map are much less marked on the $\rho$ map.

%-------------------------------------------------------------------------------
\subsection{Unified picture}
\label{section4.2}
\begin{figure}
  \centering
  \includegraphics[width=14cm]{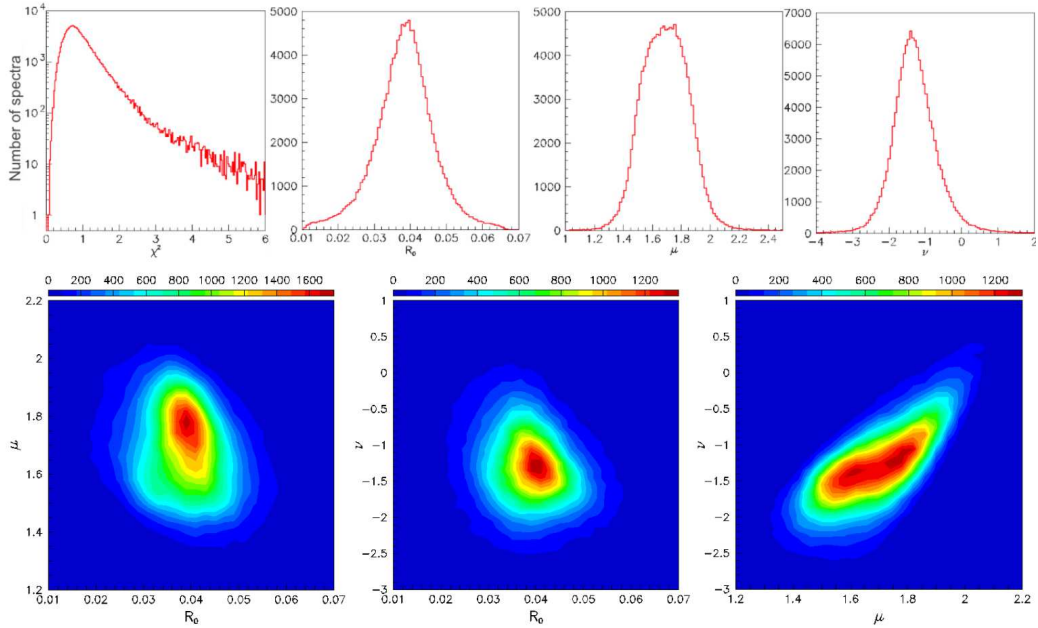}
  \caption{MASCS data. Left up: $\chi^2$ distribution of a second degree polynomial fit to the MASCS reflectance data between 400 and 800 nm. Upper right panels: distributions of the best fit parameters $R_0$, $\mu$ and $\nu$ (from left to right). Lower panels: correlation between $\mu$ and $R_0$ (left), $\nu$ and $R_0$ (central), and $\nu$ and $\mu$ (right).}
  \label{fig12}
  \end{figure}

As was done for MDIS data we fit a second degree polynomial to each measured spectrum using an uncertainty of $0.9\times10^{-3}$ on the reflectance measurements. In order to ease the comparison with MDIS data we limit the fit to the wavelength interval between 400 and 800 nm but we include all measurements in the interval, not only those corresponding to an MDIS filter wavelength as was done in the preceding Section \ref{section4.1}. Moreover, we take for $\lambda_0$ the value at the middle of the wavelength interval, 600 nm, instead of 715$\;\mathrm{nm}$. The resulting $\chi^2$ distribution (Figure \ref{fig12} left) is well behaved, with a mean value of 0.99 and a standard deviation of 0.58: no further culling of the data sample is necessary. This provides additional evidence for the large $\chi^2$ MDIS spectra to be of instrumental origin and confirms the conclusion reached from the study of the MDIS spectra: two parameters, $R_0$ and $\mu$ are sufficient to completely characterize the spectral dependence of the reflectance in the Borealis quadrangle. The mean values of the best fit parameters are listed in Table \ref{table2} for the whole Borealis region as well as for the selected regions listed in the table. Their distributions and correlations are illustrated in Figure \ref{fig12}.

\begin{figure}
  \centering
  \includegraphics[width=15cm]{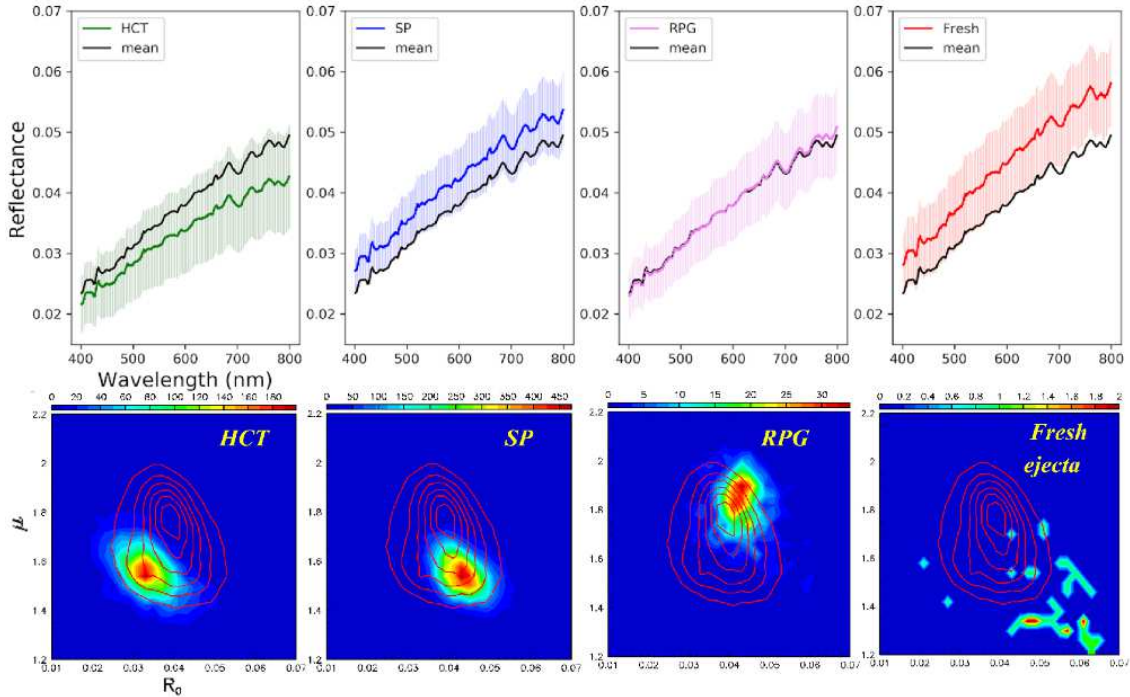}
  \caption{MASCS spectral parameters associated with the regions listed in Table \ref{table2}. Upper panels  (HCT$=$green, SP$=$blue, RPG$=$magenta, Fresh$=$red): mean spectra; error bars indicate the standard deviation and the mean spectrum associated with the whole Borealis region is shown in black. Lower panels: regions of the $\mu$ vs $R_0$ plane populated by the selected geological regions listed in Table \ref{table2}. The contours correspond to the whole Borealis region (in steps of 1 from 1 to 6 in relative units).}
  \label{fig13}
  \end{figure}

Figure \ref{fig13} displays the regions of the $\mu$ vs $R_0$ plane populated by the selected geological regions listed in Table \ref{table2}. The similarity with the results obtained for MDIS data, displayed in Figure \ref{fig5}, is remarkable. The comments that have been made in Section \ref{section3.2} can be essentially repeated here. In particular, like for MDIS data, the HCTs are seen to distort significantly the contours of the global map, introducing a split of the maximal correlation between $\mu$ and $\nu$  (Figure \ref{fig12}).    

\begin{figure}
  \centering
  \includegraphics[width=12cm]{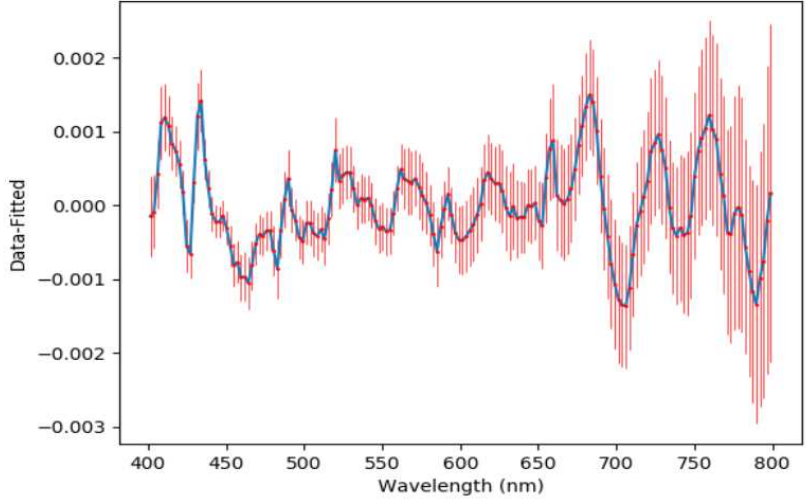}
  \caption{Distribution of the mean difference between measured and fitted MASCS reflectance. Error bars show the standard deviation of its distribution.}
  \label{fig14}
  \end{figure}

The smaller measurement uncertainty of the MASCS data compared with the MDIS data allows for a better search for possible features. Figure \ref{fig14} displays the spectrum of the difference between measured and fitted values of the reflectance. No significant outstanding feature is being revealed.

\begin{figure}
  \centering
  \includegraphics[width=15cm]{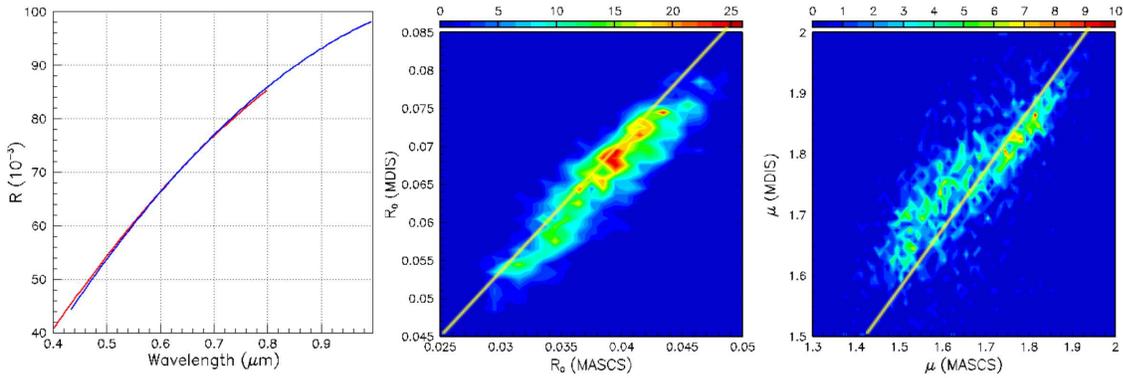}
  \caption{Comparison between the polynomial best fits to MASCS and MDIS spectra. Left: the mean polynomial best fits to MDIS (blue) and MASCS (red, scaled up by a factor 1.73) spectra. Middle and right: Correlation between the best fit parameters ($R_0$ in the middle panel and $\mu$ in the right panel) using a common reference wavelength $\lambda^{'}_{0}=\lambda_{0}=600$ nm. The yellow lines are for $R_0$(MDIS)$=$1.73$\,R_0$(MASCS) and $\mu$(MDIS)$=\mu$(MASCS)$+$0.04.}
  \label{fig15}
  \end{figure}

Finally, we compare directly the reflectance spectra described by the average polynomial best fits to the MASCS and MDIS data in the left panel of Figure \ref{fig15} after multiplication of the MASCS reflectance by a scaling factor of 1.73 (giving a better match than the factor 1.69$\pm$0.06 evaluated in Section \ref{section2.3}). The agreement is excellent. In the middle and right panels of the figure, we display the correlations between the values of the polynomial parameters using a common value of the reference wavelength, $\lambda^{’}_{0} =\lambda_0=600\;\mathrm{nm}$. The lack of a one-to-one correspondence between the footprints of MASCS measurements and the pixels of MDIS measurements prevents making a precise comparison; what is done instead is to map the values of $R_0$ and $\mu$ for MDIS and MASCS measurements separately, using square bins having a side corresponding to 1$^\circ$ in latitude; the mean values of the polynomial parameters in each bin are then used to produce the correlation plots. The correlation between the values of $R_0$ is dominated by the effect of the scaling factor. However, the correlation between the values of $\mu$ reveals a small difference, the MDIS reflectance value being $\sim$2\% (0.04) larger than the MASCS value for the bulk of the spectra, increasing up to $\sim$7\% for the larger wavelengths. These differences are too small to produce an effect visible on the left panel of the figure. Moreover, as was noted in Section \ref{section2.3}, the MDIS/MASCS ratio is known to depend on the nature of the terrain being explored and such small effects are to be expected.

%-------------------------------------------------------------------------
%-------------------------------------------------------------------------
\section{Discussion and Conclusions}
The present analysis of the spectral properties of the surface reflectance of the Borealis quadrangle are in line with what is known of the Mercury’s surface in general (\citealt{murchie2018spectral}). Its red-sloped and featureless nature, together with its low value, lower than for the Moon, is believed to be due to the presence of opaque phases, probably graphite. While lunar maria and highlands display differences in reflectance and spectral slopes caused by large variations in ferrous iron in silicates, the absence of such an effect on Mercury implies that spatial differences in spectral reflectance result instead from two major variables: variations in the content of opaque phases, which form a continuum between low and high reflectance units, and the extent of space weathering (\citealt{riner2012spectral}).

On Borealis, as on the whole Mercury surface, volcanism has played a dominant role in shaping the composition and morphology of the regolith (\citealt{head2011flood}), detailed studies of the density of craters and of the appearance of buried craters shedding light on the respective history of meteorite bombardment and of effusive volcanism. The distinction between the northern SPs and HCTs, the observation of fresh ejecta and the presence of the RPG, probably volcanic in origin, add significant information to our current understanding of Mercury’s volcanism. 

In the present work, we have analysed the spectral dependence of the surface reflectance of the Borealis quadrangle, exploiting the rich sample of observations collected by the MDIS/WAC and MASCS/VIRS instruments on-board the orbiting MESSENGER spacecraft. The former has produced a continuous and high resolution map of the reflectance at eight different wavelengths; the latter has produced continuous reflectance spectra around a large sample of different footprints. The low altitude orbit of the spacecraft when flying over Borealis allowed for a large number of images and spectra to be collected but resulted in large incidence and phase angles, causing distortions and important photometric corrections. As a result, in both cases, a significant fraction of the surface, close to the pole, could not be reliably explored. The different references used for photometry, standard for MDIS but closer to actual observing configuration for MASCS, have resulted in different scales, the former being about 1.7 times larger than the latter. Once this is taken into account, both instruments give remarkably consistent results.

In particular, both show that two parameters are sufficient to fully describe the information contained in the spectral dependence of the reflectance, one measuring the mean reflectance and the other its relative slope, or logarithmic derivative with respect to wavelength. In this context, we found it convenient to use a second degree polynomial fit to describe the measured spectra, of the form $R_{\lambda}=R_0[1+\mu(\lambda-\lambda_0)+\nu(\lambda-\lambda_0)^2]$ with $\lambda_0$ set at 715 nm for MDIS data and at 600 nm for MASCS data. We observed that over the whole Borealis quadrangle the knowledge of parameters $R_0$ and $\mu$ is sufficient to completely characterize the measured spectra, the value taken by $\nu$ being then defined to a sufficient precision. The quality of the polynomial fits was measured by the value of $\chi^2$ evaluated with measurement uncertainties of $3\times10^{-3}$ for MDIS and $0.9\times10^{-3}$ for MASCS. Apart from small deviations of low significance the second degree description gives a perfect fit to the data. Obviously, measurements performed with significantly smaller measurement uncertainties might reveal new features and invalidate our statement that two parameters are sufficient to completely characterize the spectral dependence of the reflectance.

The description in terms of two parameters, $R_0$ and $\mu$, allows for a convenient visualisation in the $\mu$ vs $R_0$ plane. The similarity between the MDIS and MASCS results is remarkable. Different units such as the SPs, HCTs, RPG and fresh ejecta occupy different regions of the plane; in particular, the global $\mu$ vs $R_0$ map reveals clearly HCTs as an entity distinct from the bulk. However, large overlaps prevent a meaningful definition of different classes on the sole basis of the properties of the spectral dependence of the reflectance.

We did not address explicitly the question of producing geologic maps of the Borealis quadrangle (e.g., \citealt{ostrach+etal+2016,ostrach2017geologic}), a topic of major interest making use of well-proven techniques, but beyond the scope of the present work. Instead, we have given a number of general information that may be of some use in getting prepared for the exploitation of the BepiColombo mission. High spatial resolution images in the 400-2000 nm range of SIMBIO-SYS on-board BepiColombo-MPO will then become available and are expected to be able to identify better the nature of regions such as the RPG and to reveal new information concerning their composition. The challenge of dealing with the large incidence angles that characterize polar observations will however remain.

%-------------------------------------------------------------------------------

%%%%%%%%%%%%%%%%%%%%%%%%%%%%%%%%%%%%%%%%%%%

\begin{acknowledgements}
This work was initiated on the occasion of a six-month stay of one of us (NBN) at the Laboratoire d’Etudes Spatiales et d’Instrumentation en Astrophysique (LESIA) de l’Observatoire de Paris, in the context of her studies at the University of Science and Technology of Hanoi (USTH). In particular the MDIS data were reduced there. We are deeply grateful to Professor Alain Doressoundiram, head of the LESIA team, who inspired this work, for his guidance and support in this early phase. Most of the subsequent analysis was then performed in the Department of Astrophysics of the Vietnam National Space Center (VNSC) of the Vietnam Academy of Science and Technology (VAST). We thank Professor Pierre Darriulat for his guidance and support in this second phase. We thank the anonymous referee for his/her useful comments which help improving the manuscript of the paper. Financial support from Centre National d'Études Spatiales (CNES), USTH is gratefully acknowledged. The data were retrieved from the PDS Geosciences Node. We thank the MESSENGER team and Océane Barraud for the reduction of the MASCS data.
\end{acknowledgements}

\label{lastpage}

\end{document}